\documentstyle[twoside,fleqn,espcrc2,epsfig]{article}

\def\be{\begin{equation}}
\def\ee{\end{equation}}
\newcommand{\beqn}{\begin{eqnarray}}
\newcommand{\eeqn}{\end{eqnarray}}

\newcommand{\AmS}{{\protect\the\textfont2
  A\kern-.1667em\lower.5ex\hbox{M}\kern-.125emS}}

\title{Heavy-Light Meson Decay Constants on the Lattice}

\author{L.Conti
        \address{Dipartimento di Fisica, Universit\`a di Roma
        ``Tor Vergata '' and INFN Sezione di Roma II, \\
        Via della Ricerca Scientifica 1, I-00133 Roma, Italy.}%
        \thanks{e-mail: livio.conti@roma2.infn.it}%
        \thanks{Work in collaboration with C.R. Allton,
M. ~Crisafulli, L. Giusti, G. Martinelli and F. Rapuano.}}
       
\begin{document}

\begin{abstract}
We present a high statistics study
of the $D$- and $B$-meson decay constants. The results were obtained
by using the Clover and Wilson lattice actions
at $\beta=6.0$ and $6.2$. 
\end{abstract}

\maketitle

\section{INTRODUCTION}
$f_B$  is a relevant parameter in the determination of the CKM matrix
elements and in the study of $B$--$\overline{B}$ mixing. 
In this paper, we present the results of a high statistics study
of the heavy-light decay constants, at $\beta=6.0$ and $\beta=6.2$, 
with the Wilson and the SW-Clover actions \cite{sw} in the quenched approximation. 
The main parameters and the details of the simulations are given
in \cite{fb}.

We extract the raw lattice value of $f_P$, $f_P^{latt}$,
using the usual ratio method
$$
f_P^{latt} \! = \! 
\left< \!
\frac{ < \!\! A_0 P^{\dag} \!\! >\!(t)}
     { < \!\! P P^{\dag} \!\! >\!(t)}
\coth(M_P(\frac{T}{2}-t)) \!
\right> \!
\frac{\sqrt{Z_{PP}}}{M_P},
$$
with $P=\overline{Q}(x) \gamma_5 q(x)$ and $A_0= \bar Q(x) \gamma_0 \gamma_5 q(x)$, where $Q$ and $q$ denote the heavy
and light quark fields respectively and $\langle...\rangle$ is a
weighted average over a given time interval 
$t_1$--$t_2$.
$M_P$ (the pseudoscalar meson mass) and $Z_{PP}$ are extracted from a fit of $< \!\! P P^{\dag} \!\! > \! (t)$ as a function of 
$t$. 	
The physical value of $f_P$ is then simply given by
\begin{equation}
f_P \equiv \frac{<\!\! 0 | A_0 | P(\vec{p}=0) \!\! >}{M_P}
      =     f_P^{latt} \; Z_{A}(a) \; a^{-1} \, .
\label{fisico}
\end{equation}
where $Z_A$ is the renormalization constant of the axial current.
Alternatively, we can extract the decay constant of the heavier mesons,
by normalizing it to $f_\pi$ (or to $f_K$), 
defined as the pseudoscalar decay constant computed 
in the chiral limit
\be 
f_P= R_P \times f_\pi^{exp}= \frac{f_P^{latt}(M_P)}{f_\pi^{latt}}
\times f_\pi^{exp} \, , \label{rp} 
\ee
where $R_P \equiv f_P^{latt}/f_\pi^{latt}$ and $f_\pi^{latt}=f_P^{latt}(M_P=0)$. 
We also introduce $R_{P_s}=f^{latt}_{P_s}/f^{latt}_K$
for the meson with the strange quark.
We have then $f_{P_s}= R_{P_s} \times f_K^{exp}$.

In order to obtain the physical values of $f_D$, $f_{D_s}$, etc.,
we have to extrapolate $f_P$ both in the heavy and light quark masses.
To be specific, we consider $R_{D_s}$, obtained 
from a linear fit in the light quark mass and (then) from  a fit in the
heavy quark mass (at fixed $m_s$) of the form 
\be f_{P_s} \sqrt{M_{P_s}}
\approx \Phi^{inf}_s + \frac{\Phi_s'}{M_{P_s}} + \frac{\Phi_s''}{M_{P_s}^2} 
+ \dots 
\, , \label{eq:phi} \ee
$\Phi^{inf}_s$, $\Phi_s'$, $\Phi''_s$ are functions which are expected to
depend
logarithmically on $m_H$  but have been taken constant in the
fit. 

{\small
\begin{table*}[bht]
\setlength{\tabcolsep}{1.5pc}
\catcode`?=\active \def?{\kern\digitwidth}
\begin{tabular*}{\textwidth}{@{}l@{\extracolsep{\fill}}ccccccc}
\hline\hline         
Run&&C60&C62&W60&W62a&W62b\\
\hline
$\beta$&&$6.0$&$6.2$ &$6.0$&$6.2$&$6.2$\\
Action && SW & SW & Wil & Wil& Wil\\
\# Confs&&170&250&120&250&110\\
Volume&&$18^3\times 64$&$24^3\times 64$&$18^3\times 64$ &$24^3\times 64$ 
&$24^3\times 64$\\
\hline
& linear& 1.56(3)& 1.48(6)&1.11(3)&1.23(4)& 1.19(5)  \\
$R_{D_s}=$&quadratic &1.57(4)&1.49(7)&1.13(4)&1.25(5)&1.20(5)\\
$f_{D_s}/f_K$&linear KLM&1.59(3)&1.50(6)&1.48(6)&1.52(5)&1.47(6)\\
&quadratic KLM&1.61(4)&1.51(7)&1.51(5)&1.55(7)&1.47(6)\\
\hline 
&linear&1.63(4)&1.58(8)&1.14(4)&1.31(6)& 1.25(7)\\
$R_{D}=$&quadratic &1.69(8)&1.73(16)&1.19(8)&1.43(11)&1.28(9)\\
$f_{D}/f_\pi$&linear KLM&1.67(4)&1.60(8)&1.52(5)&1.61(7)&1.53(8)\\
&quadratic KLM&1.72(8)&1.75(16)&1.59(10)&1.77(13)&1.59(11)\\
\hline
$f_{D_s}/f_D$&linear&1.08(1)&1.07(2)&1.06(1)&1.07(1)&1.09(2)  \\ 
&quadratic &1.09(3)&1.04(4)&1.06(3)&1.06(3)&1.13(3)\\
\hline
%
            &linear        &1.48(7) &1.28(9) &0.79(4)&0.83(4)&0.84(5)\\
$f_{B_s}/f_K$&quadratic    &1.49(9) &1.27(12)&0.81(5)&0.84(5)&0.84(5)\\
             &linear KLM   &1.56(7) &1.33(9) &1.29(4)&1.26(6)&1.16(7)\\
             &quadratic KLM&1.57(12)&1.33(11)&1.33(6)&1.27(8)&1.16(7)\\
\hline 
             &linear       &1.53(9) &1.32(13)&0.81(5) &0.86(6) &0.86(6)\\
$f_{B}/f_\pi$&quadratic    &1.57(16)&1.37(66)&0.87(9) &0.91(11)&0.86(9)\\
             &linear KLM   &1.61(10)&1.38(13)&1.32(6) &1.31(8) &1.19(8)\\
             &quadratic KLM&1.65(32)&1.43(50)&1.41(11)&1.39(15)&1.19(11)\\
\hline
$f_{B_s}/f_B$&linear       &1.10(3)&1.14(6)&1.05(2)&1.10(3)&1.12(3)\\ 
             &quadratic    &1.13(7)&1.17(17) &1.03(6)&1.14(6)&1.20(7)\\
\hline
\hline
\end{tabular*}
\caption{\it{Summary of the physical results for $R_{D_s}=f_{D_s}/f_K$,
$R_D=f_D/f_\pi$ and for 
$f_{B_s}/f_K$, $f_B/f_\pi$ obtained by extrapolating  $R_P$ and $R_{P_s}$.
We also give $f_{D_s}/f_D$ and $f_{B_s}/f_B$. 
``linear" and ``quadratic" refer to the fit in the light quark masses.}}
\label{tab:rds} 
\end{table*}
}

The major sources of uncertainty in the determination of $f_P$, 
besides the quenched approximation, come
from the calculation of $Z_A$ in eq.~(\ref{fisico}) and
from  discretization errors of $O(a)$. The use of chiral Ward identities
for a non-perturbative determination of $Z_A$
\cite{ukqcdza:mpsv}, and the ``improved"  lattice actions \cite{sw,luescher}
can help us to reduce these sources of errors.
Another method to get rid of $Z_A$ consists of extracting 
the decay constants of  heavier pseudoscalar 
mesons by multiplying $R_P$ by the experimental 
value of the pion decay constant.

Comparing the results from two different
actions we studied the reduction of the discretization errors
in the improved case and verified the validity of some KLM prescriptions \cite{lm,klm}
that have been proposed to correct $O(a)$ effects in the Wilson case.
We corrected $f_P^{latt}$ for any given
pair of values of the quark masses ($m_{1,2}$), by multiplying it by the factor

$ {\cal F}^W_{KLM}= \sqrt{(1+am_1)(1+am_2)}$ ,
in the Wilson case and by the factor \cite{clv}
$ {\cal F}^C_{KLM}= \frac{{\cal F}^W_{KLM}}{{F}_1{F}_2}$ ,
in the Clover case, where 

${F}_{1,2} = 1+\frac{1}{4} \left[
(1+a m_{1,2}) - (1+a m_{1,2})^{-1} \right]$.

\section{PHYSICAL RESULTS}
In table \ref{tab:rds} are reported the results obtained by fitting
$R_{P_s}$ ($R_P$) to eq.~(\ref{eq:phi})
both for a linear and for a quadratic fit in the light quark masses.
The scale value has been fixed from the sting tension $\sigma$ \cite{stringa}.
Although  $R_{D_s}$ is
a dimensionless quantity, the calibration  of the lattice spacing can affect
its value because it enters
in the determination of the values of the quark masses at which we extrapolate
$R_{P_s}$.
However we find the error due to the calibration of the lattice 
spacing negligible for this ratio. The same is true for the 
differents methods to fix the strange quark mass value.

Without KLM factors the results in the Wilson case
are incompatible with those obtained with the Clover action, but 
we note the remarkable
agreement between the scaled KLM-Wilson and the Clover data
at $\beta=6.0$ ($\beta=6.2$).
Within  the statistical errors 
KLM-Wilson results do not exhibit  any appreciable  $a$-dependence.
We also tested another KLM prescription \cite{BLS}, 
including the shift of the mass $M_P$, obtaining results
indistinguishable, within
the errors, from the KLM-Wilson ones reported in table \ref{tab:rds}.

In order to obtain $f_{B_s}$ and $f_B$, an extrapolation in the heavy quark mass
well outside the range available in our numerical  simulations is
needed. Discretization errors can affect the final results in two ways.
Not only do they change the actual values of the decay constants, but also
deform the  dependence of $f_P$ on $m_H$. 

\begin{figure}[hbt]
\vspace{7.5cm}
\vspace{-50cm}
\epsfbox{fd4_proc.postscript}
\epsfbox{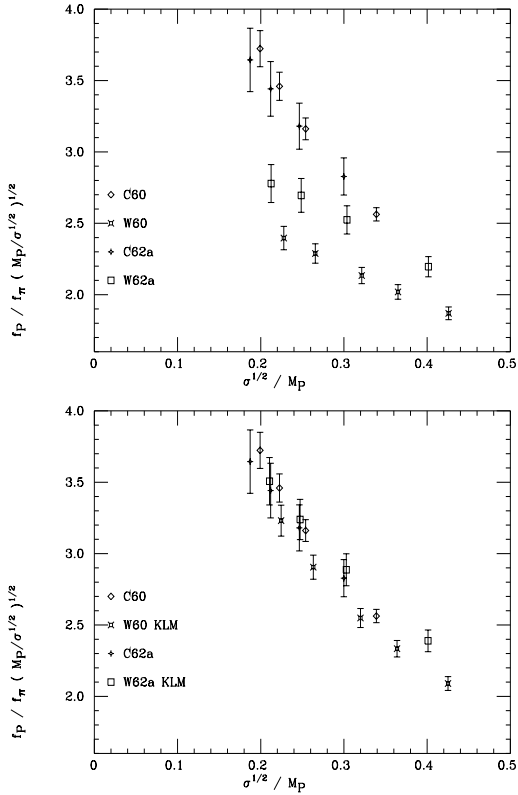}
\caption{\it{Dependence of 
$f_P/f_{\pi} (M_P/\sigma^{1/2})^{1/2}$ on $\sigma^{1/2}/M_P$.
For these points a linear extrapolation in the light quark masses
to the chiral limit has been used.}}
\label{fig:one}
\end{figure}

In fig. \ref{fig:one}, we show the Wilson and Clover results
for $f_P/f_\pi \sqrt{M_P/\sigma^{1/2}}$ as a function of the dimensionless
scale $\sigma^{1/2}/M_P$, with Wilson data uncorrected (above)
and corrected (below) by the KLM prescription:
the improvement is evident.
\vspace{7cm}
\section{CONCLUSIONS}
It is clear that, in spite of the very good 
accuracy of our data, any attempt to extrapolate
our results to $a=0$ in order to reduce discretization error would be
fruitless:
the results of the extrapolation are extremely sensitive to the choice of
the scale, given the small range in $a$ at disposal. 
Thus we believe that the best estimate
of the D- and B-meson decay constants is obtained from the Clover data at 
$\beta=6.2$, by using the method of the eq. (\ref{rp}) (from a linear fit in the light quark masses,
a quadratic fit in $1/M_P$ and without any KLM factor).
By assuming quite conservative discretization errors we found
\[\begin{array}{ll}
f_{D_s}=237 \pm 16 \; \mbox{MeV} ,& f_{D}=221 \pm 17 \; \mbox{MeV} , \nonumber \\ 
f_{B_s}=205 \pm 35 \; \mbox{MeV} ,& f_{B}=180 \pm 32 \; \mbox{MeV} , \nonumber \\ 
f_{D_s}/f_{D}=1.07 \pm 0.04   ,& f_{B_s}/f_{B}= 1.14 \pm 0.08 \, . \nonumber 
\end{array}\]
in good agreement with previous estimates \cite{flynn}.

Further studies, with comparable (or smaller) statistical
errors and physical volume,
at smaller values of the lattice spacing,  corresponding to
$\beta=6.4$ and $6.6$, are required to reduce 
the $O(a)$ dependence of the decay constants.
The use of the action of ref.~\cite{luescher} can be of great help  in this
respect. 


\begin{thebibliography}{999}
\bibitem{sw} B. Sheikloleslami and R. Wohlert, Nucl. Phys. {B259} (1985) 572;
G. Heatlie et al., Nucl.~Phys. {B352} (1991) 266.
\bibitem{fb} C.R.Allton et al., Phys.Lett. B405(1997) 133.
\bibitem{ukqcdza:mpsv} G.~Martinelli et al., Phys. Lett. B311 (1993) 241 
and Erratum Phys.~Lett. B317 (1993) 660;
D.S.~Henty et al., Phys.~Rev. D51 (1995) 5323.
\bibitem{luescher} K.~Jansen et al., Phys. Lett. { B372} (1996) 275;
M.L\"uscher et al., Nucl.Phys. {B478}(1996) 365. 
\bibitem{lm} G.P.~Lepage and P.B.~Mackenzie Phys. Rev. D48 (1993) 2250.
\bibitem{klm} A.X.~{El-Khadra} et al., Phys. Rev. D55 (1997) 3933.
\bibitem{clv} M.~Crisafulli et al., hep-lat/9707025.
\bibitem{stringa} G.~Bali and K.~Schilling, Phys.~Rev. D46 (1992) 2636. 
\bibitem{BLS} C.~Bernard, J~Labrenz and A.~Soni, Phys. Rev. D49 (1994) 2536.
\bibitem{flynn} J.~Flynn, Nucl .Phys. Proc. Suppl. 53 (1997) 168; 
G.~Martinelli, Nucl. Instrum. Meth. A384 (1996) 241
and refs. therein.
\end{thebibliography}
\end{document}